\documentclass[conference]{IEEEtran}
\usepackage{cite}
\usepackage{graphicx,epsfig,amssymb}

\ifCLASSINFOpdf
\else
\fi
\usepackage[cmex10]{amsmath}
\begin{document}
\title{Performance Analysis of Non-Orthogonal Multicast in Two-tier Heterogeneous Networks}

%\author{\IEEEauthorblockN{Xiaohu Ge$^1$, Jorge Mart\'{i}nez-Bauset$^2$, Vicente Casares-Giner$^2$, Bin Yang$^1$, Junliang Ye$^1$ and Min Chen$^3$}
%\IEEEauthorblockA{$^1$Dept. Electronics \& Information Engineering\\Huazhong University of Science \& Technology\\
%Wuhan, China\\
%xhge@mail.hust.edu.cn}
%\and
%\IEEEauthorblockN{Homer Simpson}
%\IEEEauthorblockA{Twentieth Century Fox\\
%\and
%\IEEEauthorblockN{James Kirk\\ and Montgomery Scott}
%\IEEEauthorblockA{Starfleet Academy\\
%San Francisco, California 96678-2391\\


\author{\IEEEauthorblockN{Yong Zhang\IEEEauthorrefmark{1},
Bin Yang\IEEEauthorrefmark{1},
Xiaohu Ge\IEEEauthorrefmark{1},
Yonghui Li\IEEEauthorrefmark{2}
}
\IEEEauthorblockA{\IEEEauthorrefmark{1}
School of Electronic Information and Communications, Huazhong University of Science and Technology, Wuhan, China\\}
\IEEEauthorblockA{\IEEEauthorrefmark{2}
School of Electrical and Information Engineering, University of Sydney, Sydney, Australia\\
Corresponding author: Xiaohu Ge. Email: xhge@mail.hust.edu.cn}
}
%\IEEEauthorblockA{\IEEEauthorrefmark{3}School of Computer Science \& Technology \\
%Huazhong University of Science \& Technology, Wuhan, China\\
%Email: minchen2012@mail.hust.edu.cn}}


\maketitle

\begin{abstract}
%\boldmath
With the explosive growth of mobile services, non-orthogonal broadcast/multicast transmissions can effectively improves spectrum efficiency. Nonorthogonal multiple access (NOMA) represents a paradigm shift from conventional orthogonal multiple-access (OMA) concepts and has been recognized as one of the key enabling technologies for fifth-generation (5G) mobile networks. In this paper, a two-tier heterogeneous network is studied, in which the wireless signal power is partitioned by the NOMA scheme. Moreover, the coverage probability, the average rate and the average QoE are derived to evaluate network performance. Simulation results show that compared with the OMA method, non-orthogonal broadcast/multicast method improve both the average user rate and QoE in the two-tier heterogeneous network.
\end{abstract}
\vspace{2 ex}



% \fi
\IEEEpeerreviewmaketitle

\section{Introduction}
\label{sec1}
With the explosive growth of mobile data traffic, especially video services, currently, cellular networks are facing huge challenges to provide higher spectrum efficiency for mobile users (MUs)\cite{Ge1,Ge2,Ge3,Ge4}. However, in many cases, the MU's requirements are roughly the same, e.g., requesting for hot resources. In this case, broadcast/multicast becomes a solution to achieve higher network efficiency and improve quality of experience (QoE). Multicasting enables the same content to be transmitted to all users or a specific group of users\cite{Gruber3}. Due to the growth in data traffic and the number of connected devices, traditional orthogonal multicast cannot meet the requirement of 5G multicast services at low frequencies. Nonorthogonal multiple access (NOMA) technology can achieve spectral efficiency improvement through superposition on the power domain\cite{L5}, \cite{L6}. Many studies are dedicated to NOMA's performance analysis\cite{Ding9} and energy efficiency in cellular networks\cite{Ge5},\cite{Ge6}. Compared with the traditional water-filling power allocation strategy, NOMA scheme allocates more power to users with poor channel conditions, resulting in a better compromise between system throughput and user fairness\cite{L7}. However, in the practical multicast scenario, MUs have different ability to receive the same broadcast/multicast data. Therefore, we consider a two-layer model which introduces NOMA into the network by dividing the user's service requirements into two layers, i.e., the primary layer (PL) and the secondary layer (SL). In each layer, MUs can provide the best service as they can.

Multicasting was studied in wireless networks \cite{Mirghaderi12}, heterogeneous networks (HetNets) \cite{Ding9}, and device-to-device (D2D) communications. Am¨¦rico et al. \cite{Correia13} considered scalable  MBMS video streams, with one basic layer to encode the basic quality and consecutive enhancement layers for higher quality. In this work, only the most important stream (base layer) is sent to all users in the cell. While less important streams (enhancement layers) are transmitted with less power or coding protection, only user conditions with better channels can receive additional information to improve video quality.

At the same time, since the power domain non-orthogonal transmission \cite{Dai14} enables multiple users to multiplex in the power domain, it is necessary to decode their required data from the superimposed signals through continuous interference cancellation (SIC). SIC technology can significantly increase spectrum efficiency, reduce transmission delays, and support large-scale connectivity. SIC reduces the interference power by decoding and cancelling the interference signal. A new SIC receiver was developed in \cite{Zhang15}, which decodes the signal according to the downlink signal power and subtracts the decoded signal from the received multi-user signal.
%The ordering of the received signal power depends on the transmit power of the network node, the spatial distribution of the active transmitter, and the propagation channel conditions.
However, these studies have focused on cancelling interference by NOMA schemes. How to achieve the broadcast/multicast communications by NOMA scheme is surprisingly rare in the open literature.
Utilizing the NOMA scheme in wireless signal power partitions, a two-tier heterogeneous network with NOMA scheme is proposed in this paper. The main contributions of this paper are summarized as follows:
\begin{enumerate}
\item Considering the wireless signal power partition by the NOMA scheme, a two-tier heterogeneous multicast network is proposed to provide different QoEs for MUs based on requirements and channel conditions.
\item Based on the interference cancellation scheme, the coverage probability, the average rate and the average QoE are derived for the two-tier heterogeneous multicast network with NOMA scheme.
\item Compared with the orthogonal multiple access scheme, simulation results indicate the average user rate and the QoE are improved in the two-tier heterogeneous multicast network with NOMA scheme.
\end{enumerate}

The remainder of this paper is structured as follows. Section II describes the system model. The coverage probability, the average rate and the average QoE of the two-layer heterogeneous multicast network with NOMA scheme are derived in Section III. The simulation results and discussions are presented in Section IV. Finally, conclusions are drawn in Section V.

\section{SYSTEM MODEL}

A two-tier heterogeneous network is considered in this work, where the first tier is consist of low-band macrocell base stations (MBSs) and the second tier consists of  small cell base stations (SBSs). According to \cite{Andrews16}, independently homogeneous Poisson point process (PPP) can be used to model the locations of MBSs and small cells denoted as $\Phi _M$ with density $\lambda _M$ and $\Phi _S$  with density $\lambda _S$ , respectively.

When multiple MUs request the same resource, e.g., popular video resources or news, the base station (BS) performs multicast transmission to improve transmission efficiency. As multicast MUs' channel conditions are different, in order to provide better services, the information is divided into two parts in the power domain to provide different users with different service, i.e., the primary layer (PL) which provides the basic service and the secondary layer (SL) which aims to improve the QoE of MUs. The received power from a small cell BS and a macrocell BS are denoted as $P_{rs}$ and $P_{rm}$, respectively. Besides, assume that $P_{rm}>P_{rs}$. In this context, the MU has a large probability to connect with the macrocell BS even deploying a amount of SBSs. In order to reduce the load of the MBSs, cell extension technique is adopted with a bias factor $b$ ($b > 1$). When (1) $\max \{ {P_{rs}}\}  > \max \{ {P_{rm}}\} $, MU would connect the SBS; (2) $\max \{ {P_{rm}}\}  > b\max \{ {P_{rs}}\} $, MU would connect the MBS; (3) $\max \{ {P_{rs}}\}  < \max \{ {P_{rm}}\}  < b\max \{ {P_{rs}}\} $, MU would connect the SBS. $\max \{ {P_{rs}}\}$ and $\max \{ {P_{rm}}\}$ present the received power from SBSs and MBSs, respectively. However, new problem arises that adjacent BSs may cause severe interference to the MU, e.g.,in case (3), the strongest interference power is bigger than the connected BS's power, cancellation will be applied to reduce interference. Figure 1 shows the system model and NOMA schemes for multicast communications. In the following, we will detail the channel model and interference cancellation.

\label{sec2}
 \begin{figure}
  \centering
  \includegraphics[width=8.5cm,draft=false]{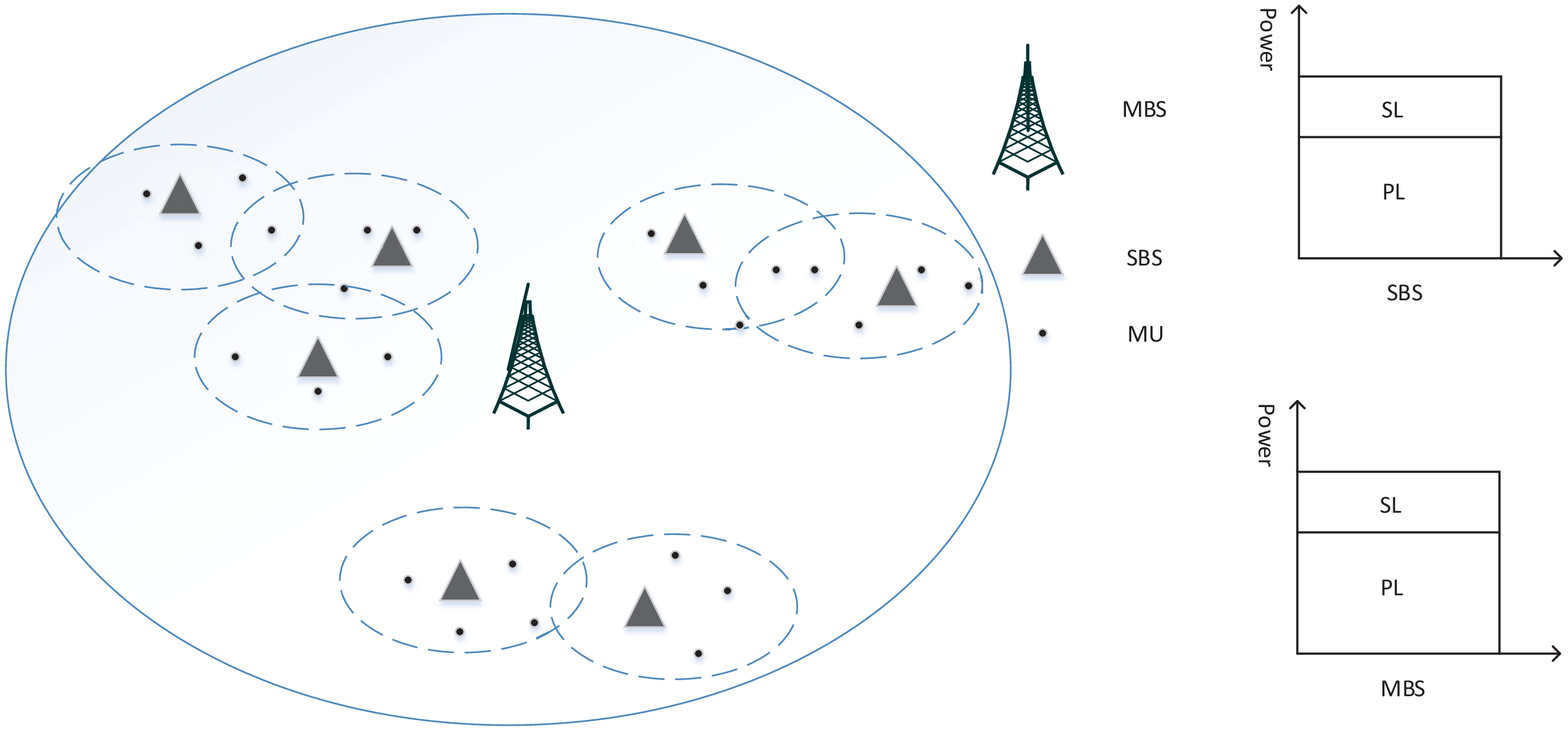}
  \caption{System model and NOMA schemes for multicast communications in two-tier heterogeneous network.}
  \label{fig1}
\end{figure}

\subsection{Path loss model}
\label{sec2-1}
To characterize shadowing effect in urban areas which is a unique scenario in our analysis, both non-line-of-sight (NLOS)  and line-of-sight (LOS) transmissions are incorporated. Specifically, given the distance between a MU and a BS, saying d, the path-loss model can be described as follows:
\begin{small}
\begin{equation}
{L_U}(d) = {C_{U,i}}{d^{ - {\alpha _{U,i}}}}\;\;,\;wp\;{P_{Ui\;}}(d),
\end{equation}
\end{small}where $U \in \left\{ {S,M} \right\}$, S and M means SBS and MBS,respectively. $i \in \{ L,N\} $, L and M represents line-of sight or non-line-of-sight, respectively. $\alpha _{U,L}$, $\alpha _{U,N}$ are the path loss exponents for BS LOS transmission and NLOS transmission respectively, $C _{U,L}$ and $C _{U,N}$  are the path loss for BS LOS and NLOS transmission at the reference distance, $ P_{UL\;}(d)$ is the probability that a link having length d is LOS, and ${P_{UN}}\;(d)\; = 1 - {P_{UL}}\;(d)$ is the probability of the NLOS one. Regarding the mathematical form of ${P_{UL\;}}(d)$, Bai[17] formulated ${P_{UL\;}}(d) = {e^{ - {\beta_U} d}}$, where ${\beta_U}$ is a parameter determined by the density and the average size of the blockages.

\subsection{Small scale fading}
\label{sec2-2}
We describe $h_i$ as the fading of the link between the i-th BS and MU. Assume that each link is subjected to Nakagami-m distribution. Then, ${H_i} = {\left| {{h_i}} \right|^2}$ follows the normalized Gamma distribution. And ${N_{SL}}$, ${N_{SN}}$, ${N_{ML}}$ and ${N_{MN}}$ are the fading parameters for the LOS link and the NLOS link in the SBS and MBS, respectively .

\subsection{Interference cancellation}
\label{sec2-2}
The interference a MU received affects decoding performance in the future 5G networks. Therefore, in order to improve the decoding capability, interference cancellation shall be applied. Assume that the useful signal is divided into two parts ${p_1} = {\alpha _p}{P_r}$ and ${p_2} = (1 - {\alpha _p}){P_r}$ by NOMA schemes for multicast communications. $p_1$ ,$p_2$ and $\alpha _p$ present the PL signal, SL signal and power allocation ratio, respectively. $X_1$, $X_2$, ..., indicate the interference signal, and without loss of generality, we assume that $X_1>X_2>...>X_k>...$ First, the MU tries to decode signal ${p_1}$ directly. If signal ${p_1}$ can't be decoded, the interference with the highest power will be decoded at the receiver. Then subtract this interference and verify whether the user can decode signal ${p_1}$ again. Due to reduce the interference cancellation complexity and latency, we assume that only one interference cancellation is performed. After decoding signal ${p_1}$, signal ${p_2}$ would be decoded by subtracting the decoded signals.

\section{Performance analysis}
\subsection{The coverage probability of the primary signal}
\label{sec2-1}
Assume that the transmission power of the SBS be $P_{ts}$, and the transmission power of the MBS be $P_{tm}$, $m=P_{tm}/P_{ts}$ ($m > 1$). As we assume that only the strongest interference is performed cancellation, $p_1$
can be successfully decoded as long as one of the following events is successful:
\begin{small}
\begin{equation}
\begin{split}
&0:{\kern 1pt} {\kern 1pt} {\kern 1pt} {\kern 1pt} {\kern 1pt} {\kern 1pt} {\kern 1pt} {\kern 1pt} {\kern 1pt} {\kern 1pt} {\kern 1pt} {\kern 1pt} {\kern 1pt} {\kern 1pt} {\kern 1pt} \frac{{{p_1}}}{{{I_{\Omega _j^0}} + {p_2} + {\sigma _S}^2}} \ge T\\
&1:{\kern 1pt} {\kern 1pt} {\kern 1pt} {\kern 1pt} {\kern 1pt} {\kern 1pt} {\kern 1pt} {\kern 1pt} {\kern 1pt} {\kern 1pt} {\kern 1pt} {\kern 1pt} {\kern 1pt} {\kern 1pt} {\kern 1pt} \underbrace {\left( {\frac{{{p_1}}}{{{I_{\Omega _j^0}} + {p_2} + {\sigma _S}^2}} < T} \right)}_{\rm{A}} \cap \underbrace {\left( {\frac{{X(1)}}{{{I_{\Omega _j^1}} + p + {\sigma _S}^2}} \ge T} \right)}_B \\
&\cap \underbrace {\left( {\frac{{{p_1}}}{{{I_{\Omega _j^1}} + {p_2} + {\sigma _S}^2}} \ge T} \right)}_C.
\end{split}
\end{equation}
\end{small}

   In order to get the coverage probability of the primary signal, three cases should be considered: (1) $\max \{ {P_{rs}}\}  > \max \{ {P_{rm}}\} $ (2) $\max \{ {P_{rm}}\}  > b\max \{ {P_{rs}}\} $ (3) $\max \{ {P_{rs}}\}  < \max \{ {P_{rm}}\}  < b\max \{ {P_{rs}}\} $. So that coverage probability can be expressed as
   \begin{small}
\begin{equation}
{P_P}({\alpha _P},{T_{PL}}) = \sum\limits_{i = 1}^3 {{P_{P,i}}({\alpha _P},{T_{PL}})},
   \end{equation}
\end{small}where ${P_{P,i}}$ is the coverage probability of the primary signal in i-th case.
\subsubsection{when $\max \{ {P_{rs}}\}  > \max \{ {P_{rm}}\} $}
\label{sec3-1}
In this case, the user is connected to the SBS and the strongest interference signal is definitely smaller than the useful signal so that interference cancellation does not need to be applied in this case ,because if the useful signal cannot be successfully decoded, the interference signal can not be decoded successfully. We define the probability of coverage in this case as
\begin{small}
\begin{equation}
\begin{split}
{P_{P,1}}({\alpha _P},{T_{PL}}) = {P_{P,1}}(Ps > Pm) = {P_{S,PL}}({\alpha _P},{T_{PL}}),
\end{split}
\end{equation}
\end{small}where
\begin{small}
\begin{equation}
\begin{split}
&{P_{S,PL}}({\alpha _P},{T_{PL}})= P(SIN{R_{PL}} > {T_{PL}}) \\
&= P(\frac{{{\alpha _p}{H_{S,0}}{L_S}({d_0})}}{{(1 - {\alpha _p}){H_{S,0}}{L_S}({d_0}) + {I_S} + {I_M} + {\sigma _S}^2}} > {T_{PL}})\\
 &= P({H_{S,0}} > \frac{{{T_{PL}}}}{{{\alpha _p} - \left( {1 - {\alpha _p}} \right){T_{PL}}}} \cdot \frac{{{I_S} + {I_M} + {\sigma _S}^2}}{{{L_S}({d_0})}})\\
 &= \sum\limits_s {{P_{S,s}}(\frac{{{T_{PL}}}}{{{\alpha _p} - \left( {1 - {\alpha _p}} \right){T_{PL}}}}),} s \in \left\{ {L,N} \right\},
\end{split}
\end{equation}
\end{small}and
\begin{small}
\begin{equation}
\begin{split}
&{I_S} = \sum\limits_{{X_s} \in {\Phi _{S,s}}\backslash {B_0}} {{C_{S,s}}{H_{S,s}}d_{S,s}^{ - {\alpha _{S,s}}} + {C_{S,\bar s}}{H_{S,\bar s}}d_{S,\bar s}^{ - {\alpha _{S,\bar s}}}} \\
&{I_M} = m\left( {\sum\limits_{{X_s} \in {\Phi _{M,i}}} {{C_{M,i}}{H_{M,i}}d_{M,i}^{ - {\alpha _{M,i}}} + {C_{M,\bar i}}{H_{M,\bar i}}d_{M,\bar i}^{ - {\alpha _{M,\bar i}}}} } \right),
\end{split}
\end{equation}
\end{small}where $s \in \left\{ {L,N} \right\},i \in \left\{ {L,N} \right\}$.
Similar with \cite{Bai17}, we can get the analytical expression as follows
\begin{small}
\begin{equation}
\begin{split}
{P_{S,s}}(T) = &\sum\limits_{n = 1}^{{N_s}} {{{( - 1)}^{n + 1}}\left( {\begin{array}{*{20}{c}}
{{N_s}}\\
n
\end{array}} \right)} \int_0^\infty  {{e^{ - \frac{{n{\eta _s}{x^{{\alpha _s}}}T\sigma _S^2}}{{{C_{S,s}}}}}}} \\
&{e^{ - {Q_{S,n}}(T,x) - {V_{S,n}}(T,x) - {Q_{M,n}}(T,x) - {V_{M,n}}(T,x)}}{f_{S,s}}(x)dx,
\end{split}
\end{equation}
\begin{equation}
\begin{split}
&{Q_{{\rm{S,}}n}}(T,x){\rm{ = }}2\pi {\lambda _S}\int_x^\infty  {F({N_{S,s}},\frac{{n{\eta _{S,s}}{x^{{\alpha _{S,s}}}}T}}{{{N_{S,s}}{t^{{\alpha _{S,s}}}}}}){p_{S,s}}(t)tdt},\\
&{Q_{M,n}}(T,x) = 2\pi m{\lambda _M}\int_{{\varphi _M}(x)}^\infty  {F({N_{M,L}},\frac{{n{\varphi _M}(x){\eta _{S,s}}T}}{{m{N_{M,L}}{t^{{\alpha _{M,L}}}}}})} {p_{M,s}}(t)tdt;\\
&{\varphi _M}(x) = {(\frac{{m{C_{M,L}}}}{{{C_{S,s}}}}{x^{{\alpha _{S,s}}}})^{1/{\alpha _{M,L}}}},\\
&{V_{S,n}}(T,x) = 2\pi {\lambda _S}\int_{{\gamma _S}(x)}^\infty  {F({N_{S,\bar s}},\frac{{n{\gamma _S}(x){\eta _{S,s}}T}}{{{N_{S,\bar s}}{t^{{\alpha _{S,\bar s}}}}}})} {p_{S,\bar s}}(t)tdt;\\
&{\gamma _S}(x) = {(\frac{{{C_{S,\bar s}}}}{{{C_{S,s}}}}{x^{{\alpha _{S,s}}}})^{1/{\alpha _{S,\bar s}}}},\\
&{V_{M,n}}(T,x){\rm{ = }}2\pi m{\lambda _M}\int_{{\xi _M}(x)}^\infty  {F({N_{M,N}},\frac{{n{\xi _M}(x){\eta _{S,s}}T}}{{{N_{M,N}}{t^{{\alpha _{M,N}}}}}}){p_{M,\bar s}}(t)tdt} ;\\
&{\xi _M}(x) = {(\frac{{m{C_{M,N}}}}{{{C_{S,s}}}}{x^{{\alpha _{S,s}}}})^{1/{\alpha _{M,N}}}},
\end{split}
\end{equation}
\end{small}
where
\begin{small}
\begin{equation}
\begin{split}
{f_{S,s}}(x){\rm{ }} =& 2\pi {\lambda _S}\exp [ - \Lambda _S^s([0,x])]\exp [ - \Lambda _M^L([0,{\varphi _M}(x)])]\cdot \\
&\exp [ - \Lambda _M^N([0,{\xi _M}(x))],
\end{split}
\end{equation}
\begin{equation}
\Lambda _{BS}^s([0,x] = 2\pi {\lambda _{BS}}\int_0^x {r{p_{BS,s}}\left( r \right)} dr,BS \in \{ S,M\} ,s \in \{ L,N\}.
\end{equation}
\end{small}

\subsubsection{when $\max \{ {P_{rm}}\}  > b\max \{ {P_{rs}}\}$}
\label{sec3-2}
The user is connected to the MBS and the strongest interference signal is definitely smaller than the useful signal in this case so that interference cancellation doesn't need to be applied in this case. We define the probability of coverage as
\begin{small}
\begin{equation}
{P_{P,2}}({\alpha _P},{T_{PL}}) = {P_{P,2}}(Pm > bPs) = {P_{M,PL}}({\alpha _P},{T_{PL}}).
\end{equation}
\end{small}
Similar with Eq. (5)
\begin{small}
\begin{equation}
{P_{M,PL}}({\alpha _P},{T_{PL}}) = \sum\limits_s {{P_{M,s}}(\frac{{{T_{PL}}}}{{{\alpha _p} - \left( {1 - {\alpha _p}} \right){T_{PL}}}})} ,s \in \{ L,N\}.
\end{equation}
\end{small}
${P_{M,s}}(\frac{{{T_{PL}}}}{{{\alpha _p} - \left( {1 - {\alpha _p}} \right){T_{PL}}}})$ could be solved the same as Eq. (7)

\subsubsection{when $\max \{ {P_{rs}}\}  < \max \{ {P_{rm}}\}  < b\max \{ {P_{rs}}\} $}
\label{sec3-3}
In this case, the user is connected to the SBS in the cell extension area, in which the maximum interference is greater than the desired signal. Therefore, cell cancellation should be adopted to improve coverage performance. We define coverage probability in this case as ${P_{P,3}}$ .
If interference cancellation is performed only once, the event of successfully decoding NOMA primary signal can be expressed as the union of the following two events.
Because event 0 is exclusive with event 1, therefore we can get the expression as follows
\begin{small}
\begin{equation}
{P_{P,3}}({\alpha _P},{T_{PL}}) = {P_{P,30}}({\alpha _P},{T_{PL}}) + {P_{P,31}}({\alpha _P},{T_{PL}}).
\end{equation}
\end{small}From Eq. (2), it is found that event 1 consists of the event A, B and C. Although the events A, B and C are related to each other which results in the difficulty to calculate ${P_{P,31}}({\alpha _P},{T_{PL}})$, we can get the approximation in some practical scenarios. Through some practical simulation we found that there is a high probability: $SIN{R_B} > SIN{R_C}$, that is, event $C \subset B$. Therefore, ${P_{P,3}}({\alpha _P},{T_{PL}})$ can be expressed as
\begin{small}
\begin{equation}
\begin{split}
{P_{P,31}}({\alpha _P},{T_{PL}}) &= P(ABC) \approx P(AC)= P(C) - P(\overline A C)\\
&= P(C) - P(\overline A )= P(A) - P(\overline C ).
\end{split}
\end{equation}
\end{small}

\vspace{0.75in}
First, we can get the expression of ${P_{P,30}}({\alpha _P},{T_{PL}}) $:
\begin{small}
\begin{equation}
\begin{split}
{P_{P,30}}({\alpha _P},{T_{PL}}) &  = {P_{P,1}}(Ps < Pm < bPs)\\
 &= {P_{P,1}}(Pm < b*Ps) - {P_{P,1}}(Pm < Ps)\\
 &= {P_{P,1}}(Ps > Pm/b) - {P_{P,1}}(Ps > Pm).
\end{split}
\end{equation}
\end{small}
Similar to Eq. (4), it¡¯s easy to calculate ${P_{P,1}}(Ps > Pm/b)$, ${P_{P,1}}(Ps > Pm)$. Second, calculating $P(A)$:$P(A) = 1 - {P_{F,30}}({\alpha _P},{T_{PL}})$. Finally, calculating $P(C)$.

In order to obtain the expression of $P(C)$, we assume that the connected link is $s \in \{ L,N\} $ and the greatest interference link is $i \in \{ L,N\} $. In this case, the connected one is SBS and the greatest interference is MBS.
\begin{small}
\begin{equation}
\begin{split}
{P_C}({\alpha _P},{T_{PL}}) &  = P(\frac{{{\alpha _p}{H_{S,0}}{L_S}({d_0})}}{{(1 - {\alpha _p}){H_{S,0}}{L_S}({d_0}) + I_S^1 + I_M^1 + {\sigma _S}^2}} > {T_{PL}})\\
&= P({H_{S,0}} > \frac{{{T_{PL}}}}{{{\alpha _p} - \left( {1 - {\alpha _p}} \right){T_{PL}}}} \cdot \frac{{I_S^1 + I_M^1 + {\sigma _S}^2}}{{{L_S}({d_0})}})\\
&= \sum\limits_{s \in \left\{ {L,N} \right\},i \in \left\{ {L,N} \right\}} {{P_{s,i}}(\frac{{{T_{PL}}}}{{{\alpha _p} - \left( {1 - {\alpha _p}} \right){T_{PL}}}})},
\end{split}
\end{equation}
\end{small}
where \begin{small}
\begin{equation}
\begin{split}
&I_S^1 = \sum\limits_{{X_s} \in {\Phi _{S,s}}\backslash B} {{C_{S,s}}{H_{S,s}}d_{S,s}^{ - {\alpha _{S,s}}} + {C_{S,\bar s}}{H_{S,\bar s}}d_{S,\bar s}^{ - {\alpha _{S,\bar s}}}} \\
&I_M^1 = m\left( {\sum\limits_{_{{X_s} \in {\Phi _M}\backslash X(1)}} {{C_{M,i}}{H_{M,i}}d_{M,i}^{ - {\alpha _{M,i}}} + {C_{M,\bar i}}{H_{M,\bar i}}d_{M,\bar i}^{ - {\alpha _{M,\bar i}}}} } \right),
\end{split}
\end{equation}
\end{small}
Referring to \cite{Bai17}, we can obtain:
\begin{small}
\begin{equation}
\begin{split}
{P_{s,i}}(T) &= \sum\limits_{n = 1}^{{N_{S,s}}} {{{( - 1)}^{n + 1}}\left( {\begin{array}{*{20}{c}}
{{N_{S,s}}}\\
n
\end{array}} \right)} \int_0^\infty  {\int_{{{(\frac{{m{C_{M,i}}}}{{b{C_{S,s}}}}x_{S,s}^{{\alpha _{S,s}}})}^{1/{\alpha _{M,i}}}}}^{{{(\frac{{m{C_{M,i}}}}{{{C_{S,s}}}}x_{S,s}^{{\alpha _{S,s}}})}^{1/{\alpha _{M,i}}}}} {} } \cdot \\
&{e^{ - \frac{{n{\eta _s}{x^{{\alpha _s}}}T\sigma _S^2}}{{{C_{S,s}}}} - {Q_{S,n}}(T,x) - {V_{S,n}}(T,x)}}\cdot\\
&{e^{- {Q_{M,n}}(T,x) - {V_{M,n}}(T,x)}}f(x,R)dRdx,\\
\end{split}
\end{equation}
\begin{equation}
\begin{split}
&{Q_{{\rm{S,}}n}}(T,x){\rm{ = }}2\pi {\lambda _S}\int_x^\infty  {F({N_{S,s}},\frac{{n{\eta _{S,s}}{x^{{\alpha _{S,s}}}}T}}{{{N_{S,s}}{t^{{\alpha _{S,s}}}}}}){p_{S,s}}(t)tdt},
\end{split}
\end{equation}
\begin{equation}
\begin{split}
{Q_{{\rm{M,}}n}}(T,x)&{\rm{ = }}2\pi {\lambda _M}\int_R^\infty  {F({N_{M,s}},\frac{{nm{C_{M,i}}{\eta _{S,s}}{x^{{\alpha _{S,s}}}}T}}{{{C_{S,s}}{N_{M,i}}{t^{{\alpha _{M,i}}}}}})}\cdot \\
 &{p_{M,i}}(t)tdt,
\end{split}
\end{equation}
\begin{equation}
\begin{split}
{V_{S,n}}(T,x)&{\rm{ = }}2\pi {\lambda _S}\int_{{{(\frac{{{C_{S,\overline s }}}}{{{C_{S,s}}}}{x^{{\alpha _{S,s}}}})}^{1/{\alpha _{S,\overline s }}}}}^\infty  \\
&{F({N_{S,\overline s }},\frac{{n{C_{S,\overline s }}{\eta _{S,s}}{x^{{\alpha _{S,s}}}}T}}{{{C_{S,s}}{N_{S,\overline s }}{t^{{\alpha _{S,\overline s }}}}}})} {p_{S,\overline s }}(t)tdt,\\
\end{split}
\end{equation}
\begin{equation}
\begin{split}
{V_{M,n}}(T,x)&{\rm{ = }}2\pi {\lambda _M}\int_{(\frac{{{C_{M,\overline i }}}}{{{C_{M,i}}}}{R^{{\alpha _{M,i}})}}^{1/{\alpha _{M,\overline i }}}}^\infty \\
 &{F({N_{M,\overline s }},\frac{{nm{C_{M,\overline i }}{\eta _{S,s}}{x^{{\alpha _{S,s}}}}T}}{{{C_{S,s}}{N_{M,\overline i }}{t^{{\alpha _{M,\overline i }}}}}}){p_{M,\overline i }}(t)tdt},
\end{split}
\end{equation}
\begin{equation}
\begin{split}
f(x,R) = &\exp [ - \Lambda _S^{\overline s }([0,{(\frac{{{C_{S,\overline s }}}}{{{C_{S,s}}}}{x^{{\alpha _{S,s}}}})^{1/{\alpha _{S,\overline s }}}}])]\\
&\exp [ - \Lambda _M^{\overline i }([0,{(\frac{{{C_{M,\overline i }}}}{{{C_{M,i}}}}{R^{{\alpha _{M,i}}}})^{1/{\alpha _{M,\overline i }}}}])]{f_{M,i}}(R){f_{S,s}}(x),
\end{split}
\end{equation}
\begin{equation}
\begin{split}
{f_{S,s}}(x) = 2\pi {\lambda _s}{p_s}(x)x \cdot \exp ( - 2\pi {\lambda _s}\int_0^x {{p_s}(t)tdt} ),\\
{f_{M,i}}(R) = 2\pi {\lambda _m}{p_m}(R)R \cdot \exp ( - 2\pi {\lambda _m}\int_0^R {{p_m}(t)tdt} ).
\end{split}
\end{equation}
\end{small}
\subsection{The coverage probability of the primary signal and the second signal}
\label{sec2-2}
\subsubsection{If the maximum interference has been successfully canceled when decoding the primary layer signal}
\label{sec3-1}
the SINR of the second layer signal can be expressed as follows:
\begin{small}
\begin{equation}
SIN{R_2} = \frac{{(1 - {\alpha _p}){H_S}{L_S}({d_0})}}{{\sum\limits_{{X_i} \in {\Phi _S}\backslash {B_0}} {{H_{S,i}}{L_S}({d_i})}  + m\sum\limits_{{X_i} \in \Phi _M^1} {{H_{M,i}}{L_S}({d_i})}  + {\sigma _S}^2}}.
\end{equation}
\end{small}

In this case, successful decoding of both signals can be represented as the following events:

\begin{small}
\begin{equation}
\begin{split}
&\underbrace {\left( {\frac{{{p_1}}}{{{I_{\Omega _j^0}} + {p_2} + {\sigma _S}^2}} < {T_{PL}}} \right)}_{\rm{A}} \cap \underbrace {\left( {\frac{{X(1)}}{{{I_{\Omega _j^1}} + p + {\sigma _S}^2}} \ge {T_{PL}}} \right)}_B \cap \\
&\underbrace {\left( {\frac{{{p_1}}}{{{I_{\Omega _j^1}} + {p_2} + {\sigma _S}^2}} \ge {T_{PL}}} \right)}_C \cap \underbrace {\left( {\frac{{{p_2}}}{{{I_{\Omega _j^1}} + {\sigma _S}^2}} \ge {T_{SL}}} \right)}_D.
\end{split}
\end{equation}
\end{small}

As already mentioned above, in general, $C \subset B$, meanwhile ${p_1}$ is bigger than ${p_2}$ because $\alpha_p$ is bigger than 0.5 and ${T_{PL}}$ is smaller than ${T_{SL}}$, therefore $SIN{R_B}/{T_{PL}} > SIN{R_D}/{T_{SL}}$, that is $D \subset B$. And we found that the value of $SIN{R_A}$ is smallest.
Also when ${\alpha _p} \le \frac{{{T_{PL}}(1 + {T_{SL}})}}{{{T_{SL}} + {T_{PL}}(1 + {T_{SL}})}}$, we can get $SIN{R_C}/{T_{PL}} > SIN{R_D}/{T_{SL}}$	
and if else, we can get $SIN{R_C}/{T_{PL}} < SIN{R_D}/{T_{SL}}$
Therefore, the successful probability can be expressed as:

\begin{small}
\begin{equation}
\begin{split}
{P_{PSL}} &= P(ABCD) = P(CD)\\
&=\left\{ {\begin{array}{*{20}{c}}
{P(A) - P(C)}&{,{\alpha _p} \le \frac{{{T_{PL}}(1 + {T_{SL}})}}{{{T_{SL}} + {T_{PL}}(1 + {T_{SL}})}}}\\
{P(A) - P(D)}&{,{\alpha _p} > \frac{{{T_{PL}}(1 + {T_{SL}})}}{{{T_{SL}} + {T_{PL}}(1 + {T_{SL}})}}}.
\end{array}} \right.
\end{split}
\end{equation}
\end{small}

Similar to the solution of $P(C)$
\begin{small}
\begin{equation}
\begin{split}
&P(D) = {P_{SL}}({\alpha _P},{T_{PL}}) = {P_{SL}}(SIN{R_2} > {T_{SL}})\\
&= {P_{SL}}({H_{S,0}} > \frac{{{T_{SL}}}}{{\left( {1 - {\alpha _p}} \right)}} \cdot \frac{{{I_{S,L}} + {I_{S,N}} + {I_{M,L}} + {I_{M,N}} + {\sigma _S}^2}}{{{L_S}({d_0})}})\\
&= {P_{SL,L}}(\frac{{{T_{PL}}}}{{1 - {\alpha _p}}}) + {P_{SL,N}}(\frac{{{T_{PL}}}}{{1 - {\alpha _p}}}).
\end{split}
\end{equation}
\end{small}

\subsubsection{If the maximum interference need not decode when decoding the first layer signal}
\label{sec3-2}
the SINR of the second layer signal can be expressed as:$SIN{R_2} = \frac{{{p_2}}}{{{I_{\Omega _j^0}} + {\sigma _S}^2}}$
In this case, successful decoding of both signals can be represented as the following events:

\begin{small}
\begin{equation}
\begin{split}
{\kern 1pt} {\kern 1pt} {\kern 1pt} \underbrace {\left( {\frac{{{p_1}}}{{{I_{\Omega _j^0}} + {p_2} + {\sigma _S}^2}} \ge {T_{PL}}} \right)}_M \cap \underbrace {\left( {\frac{{{p_2}}}{{{I_{\Omega _j^0}} + {\sigma _S}^2}} \ge {T_{SL}}} \right)}_N.
\end{split}
\end{equation}
\end{small}

Similar the successful probability can be expressed as:
\begin{small}
\begin{equation}
P(MN) = \left\{ {\begin{array}{*{20}{c}}
{P(M)}&{,{\alpha _p} \le \frac{{{T_{PL}}(1 + {T_{SL}})}}{{{T_{SL}} + {T_{PL}}(1 + {T_{SL}})}}}\\
{P(N)}&{,{\alpha _p} > \frac{{{T_{PL}}(1 + {T_{SL}})}}{{{T_{SL}} + {T_{PL}}(1 + {T_{SL}})}}},
\end{array}} \right.
\end{equation}
\end{small}
the solution of $P(M)$, $P(N)$ is similar to Eq. (5) in case 1
\subsection{The average rate of users}
\label{sec2-3}
Suppose that we can successfully decode the NOMA primary layer with threshold ${T_{PL}}$ and the NOMA second layer with threshold ${T_{SL}}$
The average rate of users can be expressed as:
\begin{small}
\begin{equation}
\begin{split}
{R_{ave}} &= \int_{{T_{PL}}}^\infty  {P(SIN{R_{PL}} = T} )\log (1 + T)dT +\\
& \int_{{T_{SL}}}^\infty  {P(SIN{R_{SL}} = T} )\log (1 + T)dT.
\end{split}
\end{equation}
\end{small}

\subsection{The average quality of experience}
\label{sec2-4}
The most widely used is the "mean Opinion Score" (MOS) proposed by the International Telecommunications Union (ITU) to evaluate the user's quality of experience (QoE). It divides the subjective perception of QoE into five levels. And according to Weber-Fechner's law, we know that the relationship between the degree of physical stimuli and its perceived intensity presents a logarithmic characteristic in many scenarios. So that we can use this property to study the evaluation of QoE\cite{Lin18}. Referring to \cite{Shao19}, the expression of MOS can be expressed in the following form:
\begin{small}
\begin{equation}
MOS(\theta ) = \left\{ {\begin{array}{*{20}{c}}
1&{,\theta  \le {\theta _1}}\\
{a\log \frac{\theta }{b}}&{,{\theta _1} < \theta  < {\theta _4}}\\
5&{,\theta  \ge {\theta _4}},
\end{array}} \right.
\end{equation}
\end{small}where $a = 3.5/\log ({\theta _4}/{\theta _1}),b = {\theta _1}{({\theta _4}/{\theta _1})^{1/3.5}}$. Thus the average service quality can be expressed as:
\begin{small}
\begin{equation}
\begin{split}
MO{S_{ave}}& = MO{S_{PL}}({P_{PL}}({\alpha _P},{T_{PL}}) - {P_{PSL}}({\alpha _P},{T_{PL}},{T_{SL}})) + \\
&MO{S_{PSL}}{P_{PSL}}({\alpha _P},{T_{PL}},{T_{SL}}).
\end{split}
\end{equation}
\end{small}And $MO{S_{PL}} = MOS({T_{PL}}),MO{S_{PSL}} = MOS({T_{PL}} + {T_{SL}})$.

\section{Results and Discussions}

In this section, the coverage probability of the primary layer, the coverage probability of both layers, the average user rate, and the average QoE under different power allocation radios are analyzed. In this simulation, some default parameters are configured with reference to \cite{Ge20},\cite{Ge21} : MBS power ${P_{tm}}=36$ dBm, SBS power ${P_{ts}}=26$ dBm, bias factor $b$ = 15, reference path loss ${C_{M,N}}=10^{-0.27}$, ${C_{M,L}}=10^{-3.08}$, ${C_{S,N}}=10^{-3.29}$, ${C_{S,L}}=10^{-4.11}$; path loss exponents: $\alpha_{M,N}=4.28$, $\alpha_{M,L}=2.42$, $\alpha_{S,N}=3.75$, $\alpha_{S,L}=2.09$; $\beta_M=0.004$, $\beta_S=0.008$; BS density $\lambda_M=10^{-5}\,BSs/{m^2}$, $\lambda_S=10^{-4}\,BSs/{m^2}$; small-scale attenuation parameter $N_{ML}=3$, $N_{MN}=2$, $N_{SL}=3$, $N_{SN}=2$; noise power ${\sigma _S} = -95$ dBm.

The effect of the coverage probability with different rate threshold is evaluated in Fig. 2. The results present that our analytical values are basically consistent with the simulation results. As the figure shows, the coverage probability of primary layer decreases gradually with the increase of $R_{pl}$. Furthermore, the more power is allocated to the primary layer, the higher coverage probability can be obtained. However, in Fig. 3, when the primary rate $R_{pl}$ = 0.1b/s/Hz, the second layer coverage probability decreases with the increase of power allocation ratio $\alpha_p$. That is because more power will be assigned to second Layer if $\alpha_p$ decreases so that the second layer signal becomes easier to be decoded, which makes it easy to get all information of both layers. The relationship between the coverage probability and the power allocation ratio in Fig. 2 and Fig. 3 is contrary. Therefore, we should balance the coverage probability of both layers, that is, balance the good channel conditions MUs' QoE and bad channel condition MUs' QoE, which is based on the fact that good channel conditions MUs can obtain both layers information while poor channel conditions MUs can only obtain the PL information.
 \begin{figure}
  \centering
  \includegraphics[width=8.5cm,draft=false]{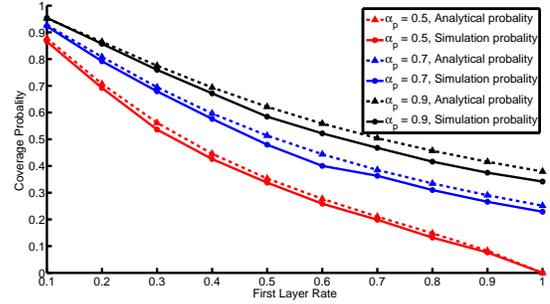}
  \caption{The analytical value and simulation value of the coverage probability of the primary layer.}
  \label{fig2}
\end{figure}

 \begin{figure}
  \centering
  \includegraphics[width=8.5cm,draft=false]{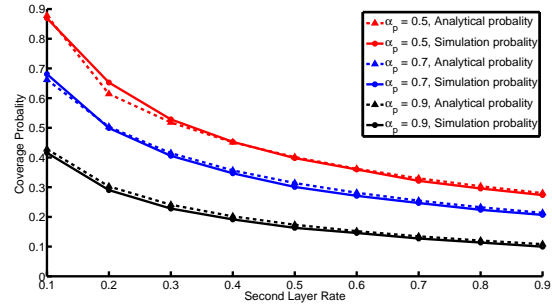}
  \caption{The analytical value and simulation value of the coverage probability of both layers when $R_{pl}=0.1$.}
  \label{fig3}
\end{figure}

 \begin{figure}
  \centering
  \includegraphics[width=8.5cm,draft=false]{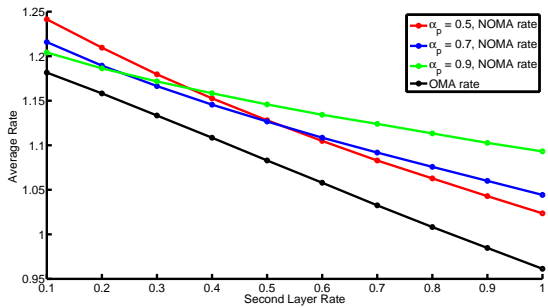}
  \caption{Average rate considering different power allocation ratios when $R_{pl}=0.1$.}
  \label{fig4}
\end{figure}

 \begin{figure}
  \centering
  \includegraphics[width=8.5cm,draft=false]{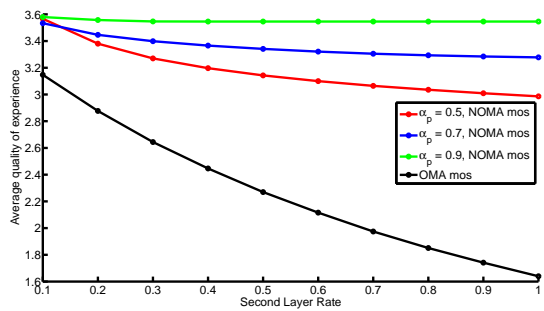}
  \caption{Average service of experience considering different power allocation ratios when $R_{pl}=0.1$.}
  \label{fig5}
\end{figure}

Fig. 4 compares the MU's average rate in NOMA and OMA considering different power allocation ratios. In OMA, the signal is transmitted as a whole and is not divided into several parts in the power domain. The results show that NOMA scheme can improve the MU's average rate, because MUs can decode signals as much as they can under different channel conditions. Strong MUs can decode all information, while weak MUs can only get basic information. As Fig. 4 shows, the average rate decreases with the increase of second layer rate, because the second layer signal is harder to be decoded successfully when the rate threshold is higher. It is also found that the curve of power allocation ratio $\alpha_p$ = 0.5 is highest when the rate of SL is less than 0.3 b/s/Hz whereas the curve of power allocation ratio $\alpha_p$ = 0.9 is highest when the rate of SL exceeds 0.4 b/s/Hz. That is because when the rate of the SL and the power allocation ratio is small, the SL can be allocated more power so the users have a greater probability to decode SL signal while the coverage probability of the PL may not reduce much. However, when the SL rate is relatively large and the power allocation ratio is small, the coverage probability of the PL is reduced but the increase of both tiers¡¯ coverage probability is not obvious.

Fig. 5 depicts QoE considering different power allocation ratios which proves that NOMA can improve the user's QoE. The green curve is always above other curves which is different from that in Fig. 4. According to the previous definition in Eq. 32, we describe the QoE with a logarithmic relationship which results in that the effect of rate on the QoE gradually decreases. Besides, the results shows that the more power is allocated to the primary layer, the QoE will be better owing to PL, which becomes easier to be decoded and basic service can be guaranteed. The increase of SL threshold rate makes it difficult to decode the SL signal while the PL signal can be easier decoded in NOMA, so the performance slightly decreased. But in OMA, it's hard to get all the information which causes bad performance for multicast MUs uses' QoE. In generally, NOMA scheme will improve multicast MUs average QoE and rate because it meets the demand of users under different channel conditions.
\section{Conclusions}

Considering the broadcast/multicast communications, a two-tier heterogeneous network with NOMA scheme is proposed in this paper. Moreover, the transmission signal power is divided into the primary layer and second layer by NOMA scheme. Furthermore, the coverage probability, average rate and average QoE are derived for a two-tier heterogeneous network. Simulation results show that proposed method can increase the quality of experience and the average rate of users for two-tier heterogeneous network with NOMA scheme. In a future work, it would be interesting to explore successive interference cancellation technology by applying for NOMA scheme in different wireless networks.

\section*{ACKNOWLEDGMENT}

The authors would like to acknowledge the support from National Key R$\&$D Program of China (2016YFE0133000): EU-China study on IoT and 5G (EXICITING-723227)





%\bibliographystyle{IEEEtran}

\end{document}